\documentclass[preprint,showpacs,showkeys,preprintnumbers,amsmath,amssymb]{revtex4}
\usepackage{graphicx}
\usepackage{dcolumn}
\usepackage{bm}
\usepackage{color}
\usepackage{bm}
\usepackage{subeqnarray}
\usepackage{pstricks}
\usepackage{pst-node}

\definecolor{darkred}{rgb}{.8,0,0}

\definecolor{darkblue}{rgb}{0,0,.7}

\def\|{|\!|}

\def\d{\textrm{d}}
\def\bol#1{\mbox{\boldmath $#1$\unboldmath}}
\newcommand{\mybox}[1]{\psshadowbox[linecolor=gray,shadowcolor=lightgray]{#1}}

\newcommand{\pw}[1]{\psframebox[linewidth=0.4pt]{\mbox{#1}}}
\newcommand{\ps}[1]{\psframebox[linewidth=0.4pt,fillcolor=lightgray,fillstyle=solid]{\mbox{#1}}}
\newcommand{\rxy}[2]{\makebox[0cm]{\raisebox{-1.6em}[0cm][0cm]{\hspace*{3mm}\mbox{#2}}}
\mybox{{\mbox{ #1}}}}

\newcommand{\ba}{\begin{array}}\newcommand{\ea}{\end{array}}
\newcommand{\be}{\begin{equation}}\newcommand{\ee}{\end{equation}}
\newcommand{\bea}{\begin{eqnarray}}\newcommand{\eea}{\end{eqnarray}}
\newcommand{\brr}{\begin{array}}\newcommand{\err}{\end{array}}
\newcommand{\bit}{\begin{itemize}}\newcommand{\eit}{\end{itemize}}
\newcommand{\ben}{\begin{enumerate}}\newcommand{\een}{\end{enumerate}}

\def\1{{_{1}}}\def\2{{_{2}}}
\def\vect#1{{\bm #1}}
\def\beq{\begin{equation}}
\def\eeq{\end{equation}}

\begin{document}

\title{Revisiting the gauge principle: enforcing constants of motion
as constraints\\}

\author{P.~Jizba}
 \email{p.jizba@fjfi.cvut.cz}
\affiliation{FNSPE, Czech Technical University in Prague,
B\v{r}ehov\'{a} 7, 115 19 Praha 1, Czech Republic\\ {\rm and} \\
ITP, Freie Universit\"{a}t Berlin, Arnimallee 14 D-14195 Berlin,
Germany\\}
\author{J.M.~Pons}
 \email{pons@ecm.ub.es}
\affiliation{Departament ECM and ICC, Facultat de F\'{\i}sica,
Universitat de Barcelona, Av. Diagonal, 647, Barcelona 08028,
Catalonia, Spain\\~\\}

\date{\today}

\begin{abstract}
In this paper we examine an alternative formulation of the gauge
principle in which the emphasis is shifted from the symmetry
transformations to their generators. We show that the gauge
principle can be entirely reformulated in terms of promoting
constants of motion - which generate rigid symmetries - to
constraints - which generate gauge symmetries. In our exposition we
first explain the basic philosophy on mechanical systems, and then
with the help of De Donder--Weyl formalism we extend our scenario
also to a field-theoretical setting. To put some flesh on bare bones
we demonstrate our method in numerous examples, including the
massive relativistic particle, the Nambu--Goto closed string and
relativistic field theory.
\end{abstract}



\maketitle

\section{Introduction}

The gauge principle (see \cite{O'Raifeartaigh} for an historical
account) is a basic ingredient of modern theoretical physics,
particularly in quantum field theory. It is not necessary to
elaborate much on this undisputable fact. A quick presentation of
its main idea is that by gauging a rigid symmetry one must pay a
``price": that of introducing a new field, the gauge field, which
geometrically represents principal connection on a
principal bundle. This ``price" has turned out to be an unexpected
bonus which has irrevocably changed the theoretical landscape in
physics.

In this  paper we propose to revisit the gauge principle from the
point of view of enforcing constants of motion as constraints.
We should, however, forewarn that our
subsequent considerations will be purely classical, so particularly
ordering issues will be outside our scope. Similarly we will assume
that Lagrangian/Hamiltonian systems are equivalent when they produce
identical equations of motion (EOM). This ``on mass-shell" (i.e.,
the classical path) identification is clearly not satisfactory from
a quantum point of view where also ``off mass-shell" behavior
non-trivially contributes into, say, transitional amplitudes.
Grassmann variables will also not be considered, since that
complication is a straightforward generalization.

It is well known that theories --- derived form a variational
principle --- which exhibit gauge invariance must be described by
constrained systems. With these two words we refer to the framework
put forward by Rosenfeld~\cite{Salisbury:2007br,rosenfeld30},
Dirac~\cite{dirac50,dirac4} and
Bergmann~\cite{bergmann49a,bergbrun49,bergm3}, who, independently,
laid the ground to deal with such systems. In particular
Rosenfeld's contribution, which has been overlooked for a long time,
has recently resurfaced thanks to the work of D. Salisbury and it is
discussed in~\cite{Salisbury:2007br}. The constrained systems are
characterized by Lagrangians whose Hessian matrix with respect to
the velocities is singular, thus preventing the Legendre map (LM)
from tangent bundle (i.e., positions and velocities
space) to cotangent  (i.e., space of positions and momenta, or phase
space) from being invertible. It is precisely the singularity of the
Hessian matrix which makes room for the possible presence of gauge
freedom. Eventually, the picture obtained in phase space is that  we
have a (non-uniquelly defined) canonical Hamiltonian -- $H$, and a
set of primary constraints $C_a$ that are just the consequence of
the non-invertibility of the LM. Thus the dynamics in phase space is
given by the Dirac Hamiltonian,
\beq H_{D} \ :=\  H + \lambda^a C_a\, , \label{h} \eeq
with $\lambda^a$ a set of in principle arbitrary Lagrange
multipliers, together with the requirement that motions must satisfy
the primary constraints,
\beq C_a\ = \ 0\,. \label{c} \eeq

Here we will not dwell in the details of the theory of constrained
systems, for which we simply refer to the
literature~\cite{Sundermeyer:1982gv,Henneaux:1992ig,Pons:2004pp,Gitman:1990}.
What we want to emphasize is that, given the structure of the
dynamics in phase space, one could think of a process of gauging a
regular theory by just starting with an ordinary Hamiltonian $H$ and
a set of functions $C_a$ that are to be enforced as constraints.
Then we could define a new dynamics by the equations (\ref{h}) and
(\ref{c}), which hopefully would describe a gauge theory.

In general, this program is bound to fail because the constraints
must have a certain degree of compatibility with the generator
$H_{\!D}$ of the dynamics. Geometrically, one needs the dynamical
trajectories to be tangent to the surface defined by the
constraints. In general, one expects this requirement to eventually
end up with the appearance of new constraints as well as the
determination of some of the Lagrange multipliers. But if $H$ and
$C_a$ are chosen too arbitrarily, the most likely outcome is that
there will be no set of $\lambda^a$'s that keeps the dynamical
trajectories tangent to the constraint's surface.

But there is a nice exception, with plenty of interest: if we choose
the would-be constraints as some of the constants of motion for $H$,
then full compatibility is easy to achieve. This is the case we will
explore. We consider a Hamiltonian for a regular theory
-- obtained from a Lagangian in tangent space through an invertible
LM -- and a set of  constants of motion $C_a$ satisfying
$\{C_a,\,H\}=0$ and closing a certain algebra $\{C_a,\,C_b\}=
 c^c_{ab} C_c$ with  $c^c_{ab}$ being
structure constants. For simplicity's sake we restrict ourselves to
constants of motion without explicit time dependent, i.e. to
scleronomic constants of motion. We then declare that the new
dynamics is governed by the Dirac Hamiltonian (\ref{h}) under the
condition that the constants of motion are enforced now as
constraints (\ref{c}).

To check that we are indeed on the right track we must verify that
with these conditions the theory defined by (\ref{h}) supports gauge
symmetries and that they act on the ``matter'' fields as they
should, just generalizing the action of the former rigid symmetries.
Once this check is done, we can explore the new gauge theory and its
dynamical consequences, because the dynamics is expected to undergo
important changes after the gauging of the rigid group of
symmetries. Finally we can further modify the theory in a natural
way by introducing gauge invariant kinetic terms for the Lagrange
multipliers. The full-fledged gauge theory is then obtained, with
the new non trivial interaction terms allowed by the gauge
principle.

Our paper is organized as follows: In Section~\ref{ngauge} we
formulate our basic strategy  using the language of mechanics.
Namely, we show how to construct a gauge invariant theory by
promoting constants of motion to constraints. We also stress an
intimate connection with the mathematical structure of non-abelian
Yang--Mills theory~\cite{Yang}. In Section~\ref{invert} we complete
the theoretical setup. Examples in mechanics are given in
Section~\ref{mechanics} and the relativistic field theory is dealt
with in Section~\ref{fields}, where the key role of the
De~Donder--Weyl formalism is made manifest. We devote
Section~\ref{NG} to the case of the closed bosonic string and use
our approach to obtain world sheet general covariance. Finally, we
conclude in Section~\ref{concl} with a brief summary of our results
and outlook.

\section{The new gauge theory}\label{ngauge}

 We start by considering a Hamiltonian for a regular theory together with a set
of scleronomic constants of motion $C_a$ satisfying $\{C_a,\,H\}=0$
and closing an algebra $\{C_a,\,C_b\}=
 c^c_{ab} C_c$.  Now we will prove that when $C_a$ are enforced now as
 constraints then this new theory is indeed a gauge theory. The simplest
way to prove it is by defining the extended Lagrangian (indices for vector
components are normally suppressed)
\beq L_{\rm e}(q,p,\dot q,\dot p,\lambda)\ = \ p\ \!\dot q - H(q,p)
- \lambda^a C_a(q,p)\,, \label{le} \eeq
and proving that it has gauge transformations. Note first that the
EOM for (\ref{le}) coincide with those derived from the Dirac
Hamiltonian (\ref{h}) and the constraints (\ref{c}) (an advantage of
the Lagrangian formulation is that all the dynamics is encoded in a
single function). Note also that we have enlarged the configuration
space with the multipliers $\lambda^a$ as new variables. We will
prove that indeed (\ref{le}) has Noether gauge symmetries. Since the
constants of motion are the Noether generators of the rigid
symmetries, it is reasonable to expect that
the generator of the  would-be canonical
gauge  transformations can be written as $G \equiv \epsilon^a(t) C_a$,
with $\epsilon^a$ being a set of arbitrary time-dependent functions.
We will prove now that indeed $G$ generates gauge transformations.
The corresponding variations can be written as
\beq \delta_{\epsilon} q^i\ = \  \{q^i,G\}, \;\;\; \delta_{\epsilon}
p_i\  = \ \{p_i,G\}\, , \label{deltapq} \eeq
and the variations of the multipliers will be determined below by the
condition that, under the variations thus defined, the Lagrangian
$L_{\!\rm e}$ is quasi-invariant, i.e.,
\beq \delta_{\epsilon}L_{\rm e} \ = \ \frac{\d}{\d t} F\, , \eeq
for some $F$ linear in $\epsilon$ and its
derivatives. Indeed,
\begin{eqnarray}
\delta_{\epsilon} (p_i\ \! \dot q^i) \ &=& \ p_i \delta_{\epsilon} \dot
q^i \ + \ \dot q^i \delta_{\epsilon} p_i \ = \ - \dot p_i
\delta_{\epsilon} q^i \ + \ \dot q^i \delta_{\epsilon} p_i \ + \
\frac{\d}{\d t} \!\left( p_i \delta_{\epsilon} q^i \right) \nonumber
\\[2mm]
&=& \ - \epsilon^a(t) \frac{\partial C_a}{\partial p_i} \dot p_i  \ - \
\epsilon^a(t) \frac{\partial C_a}{\partial q^i} \dot q^i \ + \
\frac{\d}{\d t}\!\left( p_i \epsilon^a(t) \frac{\partial C_a}{\partial
p_i}
\right)\nonumber \\[2mm]
&=& \ \dot \epsilon^a(t) C_a \ + \ \frac{\d}{\d t}
\left[\epsilon^a(t)\left(p_i \frac{\partial C_a}{\partial p_i} - C_a
\right) \right]\! ,
\end{eqnarray}
and
\begin{eqnarray}
\delta_{\epsilon}(H + \lambda^a C_a)  =  \epsilon^b(t) \{H, C_b\} +
C_a \delta_{\epsilon}\lambda^a  +  \lambda^a \epsilon^b(t)
\{C_a,C_b\}  = C_a \delta_{\epsilon}\lambda^a  +  \lambda^a
\epsilon^b(t)  c_{ab}^c C_c\, .
\end{eqnarray}
Thus, the appropriate definition
\beq \delta_{\epsilon}\lambda^a \ := \  \dot \epsilon^a(t) -
\lambda^b \epsilon^c(t)  c_{bc}^a \  =: \ ({{D}}_0\, \epsilon(t))^a\,
, \label{deltalambda} \eeq
makes $\delta_{\epsilon} L_{\rm e}$ to be
\begin{eqnarray}
\delta_{\epsilon} L_{\rm e}  &=&  C_a \left(\dot \epsilon^a(t) -
\delta_{\epsilon}\lambda^a - \lambda^b \epsilon^c(t) c_{bc}^a
\right) +  \frac{\d}{\d t} \left[\epsilon^a(t)\left(p_i
\frac{\partial C_a}{\partial p_i} - C_a \right) \right]\nonumber
\\[2mm]
&=& \frac{\d}{\d t} \left[\epsilon^a(t)\left(p_i \frac{\partial
C_a}{\partial p_i} - C_a \right) \right] \! ,
\end{eqnarray}
which proves that the variations (\ref{deltapq}) and
(\ref{deltalambda}) define a Noether gauge symmetry for $L_{\rm e}$.
In Eq.(\ref{deltalambda}) we have introduced the covariant
derivative \cite{Gomis:1993ki}
\begin{eqnarray}
({{D}}_0)_c^a\ := \ \partial_t \delta^a_c\ - \ \lambda^b
c_{bc}^a\, , \label{2.10aa}
\end{eqnarray}
which is nothing but the covariant derivative for the adjoint
representation. Analogously, one can introduce the covariant
derivative for the phase-space variables $\xi^i = \{p_1,
\ldots, q^1, \ldots\}$
as
 \begin{eqnarray}\label{2.11ab}
 {{D}}_0  \xi^i \ := \  \partial_t {\xi^i} \ - \ \lambda^a  \Gamma_a\ \!{\xi^i}  \ = \
  \partial_t {\xi^i} \ - \ \lambda^a \{ { \xi^i},\, C_a\}
\end{eqnarray}
Here  $\Gamma_a$ is the  representation of the symmetry generators
$C_a$ that acts on  $\xi^i$. Indeed, the
Jacobi identity for Poisson brackets ensures that  $[\Gamma_a,
\Gamma_b] = -c_{ab}^c\Gamma_c$.  Because both $\partial_t$
and $\{~, C_a\}$ fulfil the Leibniz rule one can extend the
covariant derivative (\ref{2.11ab}) to any function
$\phi(\vect{\xi})$ on phase space.

Using the fact that our active variations commute with the time
derivatives,  i.e., $\delta_\epsilon (\partial_t
 \phi) =
\partial_t(\delta_\epsilon  \phi)$, it is easy to check that the
covariance condition takes the form
\begin{eqnarray}
\delta_\epsilon (D_0 \,\phi) = \epsilon^a D_0 (\{\phi,\,C_a\}) =:
 \epsilon^a \Gamma_a (D_0 \, \phi)\, . \label{2.12}
\end{eqnarray}

The first equality in (\ref{2.12}) can be proved by considering: 1)
that active variations commute with the time derivatives, and so
$\Gamma_a (\partial_t { \phi})= \partial_t\{ {\phi},\, C_a\}$, and
2) that the action of $\Gamma_a$ on the multipliers is the adjoint
action: $\Gamma_a \lambda^b = c^b_{ac}\lambda^c$. Notice that in the
last equality in (\ref{2.12}) the representation $\Gamma_a$ of the
symmetry generators acting on $D_0\, q$ was defined. This definition
turns out to be an exact identity on mass-shell. In this regard it
is interesting to realize that the curvature ${\bf F} =
[{{D}}_0,{{D}}_0] = 0$, and  so in the case of mechanics the usual
gauge-invariant kinetic term $\mbox{Tr}({\bf F}^2)$  is trivially
zero. Thus the multipliers $\lambda^a$  cannot become dynamical
variables. On the other hand, a subsequent elimination of the
momenta --- which are auxiliary variables (auxiliary variables are
by definition variables that can be isolated by using their own EOM)
for $L_{\rm e}$ ---, as done in the next subsection, will assign the
$\lambda^a$'s the status of auxiliary variables.

Note that (\ref{deltalambda}),  (\ref{2.10aa})
 and (\ref{2.11ab}) carry indeed all
the flavor of the transformation of a gauge field in a non-abelian
gauge theory. This is exactly the case, because what we have done is
precisely the application of the gauge principle: to gauge a group
of rigid symmetries.  We remind that the rigid symmetries are
generated by constants of motion while the gauge symmetries by the
first class constraints~\cite{foot4}. Thus  gauging a group of rigid
symmetries is tantamount to enforce the generating constants of
motion as constraints. In this respect $\lambda^a$ play the role of
a connection in a principal bundle over ${\mathbb{R}}$. The fact
that ${\bf F} =0$ then indicates that this bundle is flat (not big
surprise for a bundle with so simple base space). Let us, however,
stress that our derivation would go through even if not {\revised
every} rigid symmetry is gauged. For instance, we could have limited
ourselves only to gauging any subgroup of rigid symmetries. The
analogy with non-abelian Yang--Mills theory is summarized in
Table~\ref{tab1}.
\begin{table}
\begin{ruledtabular}
\caption{\label{tab1}\footnotesize{Comparison between the gauge
theory presented in Section~\ref{ngauge} and the non-abelian
Yang--Mills theory. The parallelism obtained allows to formally identify
$D_0 \leftrightarrow D_{\mu}$ and ${\vect{\lambda} \leftrightarrow
{\vect{A_{\mu}}} }$.}\\}
\begin{tabular}{ll}
Gauge theory from Section~\ref{ngauge}\footnote{\scriptsize {Here we
accept notations: $\vect{\lambda} = \lambda^aC_a$, $\vect{\epsilon}
= \epsilon^a C_a$, $\xi = \{p_1,p_2, \ldots, q^1,q^2, \ldots\}$ is a
phase-space point, ${\phi}$ is an arbitrary function on a phase space
and $\Gamma(C_a) = \Gamma_a = \{~,
C_a\} = \omega^{ij}\frac{\partial C_a}{\partial \xi^j} \frac{\partial}{\partial
\xi^i}$. }\\[-3mm]}
& Non-abelian Yang--Mills theory\footnote{\scriptsize{Here we
accept notations: $\vect{A}_{\mu} = -i A^a_{\mu}t_a$, $\vect{\epsilon} =
\epsilon^at_a$, $\vect{\Phi}$ is an arbitrary field multiplet and
$T(t_a)  = T_a$ is an irreducible
representation of the algebra of $t_a$ generators that is adapted to $\vect{\Phi}$, e.g. for $\vect{\Phi}$ in fundamental rep. of $SU(N)$ then $T({\vect A}_{\mu}) = -iA^a_{\mu} T_a$ with $T_a$ being the $(N\times N)$ hermitian matrices. Generators in self-adjoint rep. are defined as, $(T_b)_c^a = i f_{bc}^a$. } }\\
\hline
\footnotesize{$\delta_{\epsilon}\lambda^a(t) = D_0\, \epsilon^a(t)$} &
\footnotesize{$\delta A^a_{\mu}(x)= D_{\mu}\,\epsilon^a(x)$} \\
\footnotesize{$\delta_{\epsilon}\phi(\xi) = \Gamma(\vect{\epsilon}) \phi(\xi)$} &
\footnotesize{$\delta \vect{\Phi}(x) = iT(\vect{\epsilon})\vect{\Phi}(x) $}\\
\footnotesize{$D_0\,\epsilon^a(t) = \partial_t\epsilon^a(t) - \lambda^b(t)
c^a_{bc} \ \!\epsilon^c(t)$}
&
\footnotesize{$D_{\mu}\,\epsilon^a(x) = \partial_{\mu}\epsilon^a(x) + A^b_{\mu}(x)f^a_{bc}\
\!\epsilon^c(x)$} \\
\footnotesize{$D_0\,\phi(\xi) = \partial_t \phi(\xi) -
\lambda^a \Gamma_a\phi(\xi)$}
&
\footnotesize{$D_{\mu}\vect{\Phi}(x)=
\partial_{\mu}\vect{\Phi}(x)  -i A_{\mu}^a T_a\vect{\Phi}(x)$}
\\
\footnotesize{$\{C_a,C_b\} = c_{ab}^c C_c \ \Rightarrow \
[\Gamma(C_a),\Gamma(C_b)] = - c_{ab}^a\Gamma(C_a)$} &
\footnotesize{$[t_a,t_b] = if_{ab}^c t_{c} \ \Rightarrow \ [T(t_a),
T(t_b)] = if_{ab}^c
T(t_{c})$}\\
\hline \hline On-mass-shell situation ($\partial_tC_a = 0$)\\
\hline \footnotesize{$D_0\,\vect{\epsilon} =
\partial_t\vect{\epsilon} - \{ \vect{\lambda},\vect{\epsilon} \}$ }
&
\footnotesize{$D_{\mu}\,\vect{\epsilon} =
\partial_{\mu}\vect{\epsilon} + [{\vect{A}}_{\mu},\vect{\epsilon}]$}
\\
\footnotesize{$\delta_{\epsilon} (D_0\, \phi) = D_0(\{\phi, \vect{\epsilon} \}) =
\Gamma(\vect{\epsilon})D_0\, \phi$} &
\footnotesize{$\delta (D_{\mu}\, \vect{\Phi}) = iT({\vect{\epsilon}})D_{\mu}\,\vect{\Phi}$}\\
\end{tabular}
\end{ruledtabular}
%
\end{table}

Note finally that the case of a soft algebra~\cite{Henneaux:1992ig} is
easily accommodated.
We can relax the condition that the constants of motion $C_a$ form a
Lie algebra to that of a soft algebra, where there are no longer
structure constants but structure functions, $\{C_a,\,C_b\}=
 c^c_{ab}(q,p) C_c$, and we can also relax the constant of motion
condition, $\{C_a,\,H\}=0$, to $\{C_a,\,H\}=a_a^b(q,p) C_b$. In this
case, equation (\ref{deltalambda}) changes to
\beq \delta_{\epsilon}\lambda^a \ := \ \dot \epsilon^a(t)\ - \ \lambda^b
\epsilon^c(t)  c_{bc}^a \ - \ \epsilon^b(t)a_a^b\, . \label{deltalambda2}
\eeq

In field theory one can find more general cases
\cite{Henneaux:1992ig}, like that of an open algebra, where the
algebra of the constants of motion only closes up to linear terms
that are antisymmetric combinations of the equations of motion, or
when there is functional dependence among the constants of motion.
We believe that these cases can also be addressed, but since the
ordinary case already requires a non-standard formalism (see
Subsection~\ref{dd}) we leave them for further study.

\section{Inverting the Legendre map}\label{invert}

\subsection{The Lagrangian $L_{\!\lambda}$}

Consider now the equations of motion for $L_{\rm e}$,  i.e.,
\begin{subeqnarray}\label{eom1} &&\dot q \ - \ \frac{\partial H}{\partial p} \ - \ \lambda^a
\frac{\partial C_a}{\partial p} \ = \ 0 \,
, \\[1mm]
&&\dot p \ + \ \frac{\partial H}{\partial q} \ + \ \lambda^a
\frac{\partial C_a}{\partial q} \ = \ 0 \, ,
 \\[1mm]
&&C_a \ = \ 0\, .
\end{subeqnarray}
It is interesting to observe that by introducing the
symplectic matrix ${\bol{\omega}}$
\begin{eqnarray}
{{\omega}}_{ij} ~= ~\left(\begin{array}{cc} 0&{\openone}~\\
-{\openone}~&0
\end{array}\right)_{ij}\, ,
\end{eqnarray}
(with ${{\omega}}_{ij}^{-1} =  {\omega}^{ij}$) the EOM (\ref{eom1})
can be succinctly written as
\begin{eqnarray}
D_0 \xi^i \ = \  {\omega}^{ij}\ \!\frac{\partial H}{\partial
\xi^j}\, , \ \ \ \ \ \ C_a \ \ = \ 0\, .
\end{eqnarray}

We can now use the first set of Eq.(\ref{eom1}) to isolate the
momenta in terms of positions $q$, velocities $\dot q$ and the
multipliers $\lambda^a$, thus rewriting (\ref{eom1}a) in the
equivalent form
\beq p \ - \ P(q,\dot q,\lambda)\ = \ 0\,, \label{eom1bis} \eeq
for some functions $P$. This invertibility of the LM will hold in
general. In fact, since the starting theory was not
gauge, invertibility is guaranteed for $\lambda^a=0$. With
$\lambda^a$ being just new independent variables, invertibility will
be maintained in general.

We implement $p\to P(q,\dot q,\lambda)$ into $L_{\rm e}$ to define
the new Lagrangian $L_{\!\lambda}$,
\begin{eqnarray} L_{\!\lambda}(q,\dot q,\lambda)\ = \ P(q,\dot
q,\lambda)\dot q \ - \ H(q,P(q,\dot q,\lambda)) \ - \ \lambda^a
C_a(q,P(q,\dot q,\lambda))\,.
\end{eqnarray}
Notice then
\begin{eqnarray}
\frac{\partial L_{\!\lambda}}{\partial q} \ &=& \
-\left.\Big(\frac{\partial H}{\partial q} \ + \
\lambda^a\frac{\partial C_a}{\partial q}\Big)\!\right|_{{p\to P}}\ +
\ \left.\Big(\dot q- \frac{\partial H}{\partial
p}-\lambda^a\frac{\partial C_a}{\partial p}\Big)\!\right|_{{p\to
P}}\ \! \frac{\partial P}{\partial \dot
q}\nonumber \\[1mm]
&=& \ -\left.\Big(\frac{\partial H}{\partial q} \ + \
\lambda^a\frac{\partial C_a}{\partial q}\Big)\!\right|_{{p\to P}}\,
,
\end{eqnarray}
because $\frac{\partial L_{\rm e}}{\partial p}|_{{p\to P}}=
\left.\Big(\dot q- \frac{\partial H}{\partial
p}-\lambda^a\frac{\partial C_a}{\partial p}\Big)\!\right|_{{p\to
P}}$ vanishes identically owing to the procedure to define the
functions $P(q,\dot q,\lambda)$. By the same token, we obtain
\begin{eqnarray}
\frac{\partial L_{\!\lambda}}{\partial \dot q} \ = \  P(q,\dot
q,\lambda)\, , \label{eom4}
\end{eqnarray}
so we reobtain the functions $P$ as the definition of the new
Lagrangian momenta. By taking into account Eq.(\ref{eom4}) the reader may note
that the EOM for the Lagrangian $L_{\!\lambda}$ yield
\beq \frac{\d P}{\d t} \ + \ \frac{\partial H}{\partial q} \ + \
\lambda^a\frac{\partial C_a}{\partial q} \ = \ 0\, ,  \label{eom2bis}
\eeq
 which  is equivalent to
(\ref{eom1}b) when the identity (\ref{eom1bis}) is utilized.

The remaining EOM for $L_{\rm e}$ is the one associated
with the multiplier $\lambda$. This equation sets just the
constraint as EOM. From the perspective of
$L_{\lambda}$, we can write down EOM for $\lambda^a$;
$\frac{\partial L_{\!\lambda}}{\partial \lambda^a} = \frac{\partial
L_{\rm e}}{\partial p}|_{_{p\to P}}\ \!\frac{\partial P}{\partial
\lambda^a} - C_a(q,P(q,\dot q,\lambda))$, but again, since
$\frac{\partial L_{\rm e}}{\partial p}|_{_{p\to P}}$ vanishes
identically, we end up with
\beq C_a(q,P(q,\dot q,\lambda))\ = \ 0\, ,
\label{eom3bis}
\eeq
as the last EOM for $L_{\!\lambda}$. This shows the equivalence
between EOM from $L_{\rm e}$ and $L_{\!\lambda}$, because
(\ref{eom3bis}) is just (\ref{eom1}c) with the substitution $p\to
P(q,\dot q,\lambda)$, which is nothing but the EOM (\ref{eom1}a).

\subsection{Gauge symmetry for $L_{\!\lambda}$}

Let us now prove that $L_{\!\lambda}$ has the gauge symmetry
$\delta_{\lambda} q = (\delta_{\epsilon} q)|_{p\to P},\
\delta_{\lambda} \lambda = \delta_{\epsilon}\lambda$. One has
\begin{eqnarray}
\delta_{\lambda} L_{\lambda}\ = \ \left.\left(\frac{\partial L_{\rm
e}}{\partial q}\right)\!\right|_{{p\to P}}\!\! \delta_{\lambda} q \
+ \ \left.\left(\frac{\partial L_{\rm e}}{\partial \dot
q}\right)\!\right|_{{p\to P}}\!\! \delta_{\lambda} \dot q  \ + \
\left.\left(\frac{\partial L_{\rm e}}{\partial
p}\right)\!\right|_{{p\to P}}\!\! \delta_{\lambda} P \ + \ \left.
\left(\frac{\partial L_{\rm e}}{\partial
\lambda}\right)\!\right|_{{p\to P}}\!\! \delta_{\lambda} \lambda\, .
\end{eqnarray}
We need not care of the term with $\delta_{\lambda} P$ because
$(\frac{\partial L_{\rm e}}{\partial p})|_{p\to P}=0$ identically
due to the equivalence between (\ref{eom1}) and (\ref{eom1bis}). As
regards $\delta_{\lambda} \dot q$ we can write it as
$(\delta_{\epsilon} \dot q)|_{p\to P}$. All in all we can write
\begin{eqnarray}
\delta_{\lambda} L_{\!\lambda}\ = \ (\delta_{\epsilon} L_{\rm
e})|_{p\to P} \ = \ \left.\left(\frac{\d}{\d t}
F\right)\!\right|_{p\to P}\ = \ \frac{\d}{\d t}( F_{\ \!|_{p\to
P}})\, ,
\end{eqnarray}
where~\cite{foot7} in the last step we use that $p\to P$ implies
also $\dot p\to \frac{\d}{\d t} P$, etc. Thus we have proved that
$L_{\!\lambda}$ inherits the gauge invariance  of $L_{\rm e}$.

\subsection{A step further}

 Finally, if $\lambda$ can be isolated from the equation
(\ref{eom3bis}), this means that it is in fact an auxiliary
variable. It is well known that auxiliary variables can be
substituted back into the Lagrangian without affecting the dynamics
(see e.g., the Appendix in \cite{Pons:1999az}). In fact
the earlier substitution $p\to P(q,\dot q,\lambda)$ in the previous
subsection is an example of this mechanism, for the variables $p$
are isolated by use of their own equations of motion (\ref{eom1}),
but we have been explicit in the proof of equivalence of EOM. Thus
with the substitution we would have arrived at a new Lagrangian
$L(q,\dot q)$ with a dynamics equivalent to that of $L_{\rm e}$. Of
course, there may be technical obstacles to carrying out this step:
solving the system of equations (\ref{eom3bis}) may prove too
difficult, getting rid of the multipliers can lead in general to
impractically complicated, non polynomial expressions for $L$, etc.
One can then revert back to the Lagrangian $L_{\rm e}$, with its EOM
(\ref{eom1}).

\subsection{$\cdots$ and a step further}\label{IIID}

Despite potential complications
related to solving the system (\ref{eom3bis})  we will suppose that
indeed the variables $\lambda^a$ can be isolated from the equations
(\ref{eom3bis}) and eliminated by plugging them back
into the Lagrangian  $L_{\!\lambda}$. Thus (\ref{eom3bis})
will be equivalent to $\lambda^a = \Lambda^a(q,\dot q)$ for some
functions $\Lambda^a$. We will prove that $L(q,\dot q) :=
(L_{\!\lambda}(q,\dot q,\lambda))|_{\lambda\to \Lambda}$ has the gauge
symmetry $\delta_{L} q = (\delta_{\lambda} q)|_{\lambda\to
\Lambda}$.  One has
\begin{eqnarray}
\delta_L L \ = \ \left.\left(\frac{\partial L_{\!\lambda}}{\partial
q}\right)\!\right|_{\lambda\to \Lambda}\!\! \delta_{L} q \ + \
\left.\left(\frac{\partial L_{\!\lambda}}{\partial \dot q}\right)\!\right|_{\lambda\to
\Lambda}\!\! \delta_{L} \dot q \ + \ \left.\left(\frac{\partial
L_{\lambda}}{\partial \lambda}\right)\!\right|_{\lambda\to \Lambda}\!\! \delta_{L}
\lambda\,.
\end{eqnarray}
Note that we do not have to define $\delta_{L} \lambda$ because the equation $\lambda =
\Lambda(q,\dot q)$ is exactly $\frac{\partial L_{\lambda}}{\partial \lambda}=0$. We continue
\begin{eqnarray}
\mbox{\hspace{-3mm}}\delta_L L \ = \ \left.\left(\frac{\partial
L_{\lambda}}{\partial q}\delta_{\lambda} q + \frac{\partial
L_{\lambda}}{\partial \dot q}\delta_{\lambda} \dot q  +
\frac{\partial L_{\lambda}}{\partial \lambda}\delta_{\lambda}
\lambda\right)\!\right|_{\lambda\to \Lambda} \! = \
\left.\left(\frac{\d}{\d t}F_{\ \!|_{p\to
P}}\right)\!\right|_{\lambda\to \Lambda} \! = \ \frac{\d}{\d
t}\left(F_{\ \!|_{p\to P,\ \lambda\to \Lambda}}\right)\!.
\end{eqnarray}
This concludes the proof that $L$ is a Lagrangian with gauge
symmetry. Our result is general. Given any regular (i.e., non-gauge)
theory and a Noether constant of motion in the canonical formalism,
one can make this constant of motion a first class constraint and
construct an associated Lagrangian with this gauge symmetry.

\section{Examples in mechanics}\label{mechanics}

\subsection{Enforcing a function not being a constant of motion as
a constraint}

Although we are developing the theory for implementing constants of
motion as constraints, let us consider an example where one
implements a non-constant of motion, just to realize in practical
terms the problems that are likely to appear. Consider the standard
Hamiltonian $H({\vect q},{\vect p}) = \frac{{\vect p}^2}{2\,m} +
V({\vect q}^2)$ (${\vect q}$ and ${\vect p}$ are $d$-dimensional
vectors) and try to implement $C({\vect q},{\vect p}) = {\vect
q}\cdot{\vect p}$ as a constraint. Following the above instructions
we get ${\vect P}({\vect q},\dot {\vect q},\lambda) = m (\dot{\vect
q}-\lambda\vect q)$ and $\lambda$ is determined as $\Lambda({\vect
q},\dot {\vect q})= m\frac{{\vect q}\cdot\dot{\vect q}}{{\vect
q}^2}$. A substitution of both determinations of $\vect p$ and
$\lambda$ into the extended Lagrangian yields $L({\vect q},\dot
{\vect q}) = \frac{1}{2}m \,\dot {\vect q}\ \! \mathbb{M}\ \!  \dot
{\vect q} - V(\vect q^2)$, where $\mathbb{M}$ is the matrix
$\mathbb{M}_{ij}= \delta_{ij} - \frac{q_i\,q_j}{{\vect q}^2}$. This
Lagrangian is singular because the Hessian matrix with respect to
the velocities, is (up to a multiplicative constant) identical to
$\mathbb{M}$, i.e. to a projector transverse to ${\vect q}$. Thus
$L({\vect q},\dot {\vect q})$ may potentially describe a gauge
theory.

The Lagrangian momenta are defined as $\hat {\vect p} =
\frac{\partial L}{\partial \dot {\vect q}} = \mathbb{M}  \dot {\vect
q}$, which indeed implies the constraint ${\vect q}\cdot{\vect p}
\simeq 0$ because $ \mathbb{M}{\vect q} = 0$ identically. The
canonical Hamiltonian is just $\frac{{\vect p}^2}{2\,m} + V({\vect
q}^2)$. So the dynamics in phase space is given by the Dirac
hamiltonian $H_{D}({\vect q},{\vect p}):= H({\vect q},{\vect p}) +
\eta\ \!{\vect q}\cdot{\vect p}$, as expected. The problem in this
example is that we must require stabilization of the now primary
constraint ${\vect q}\cdot{\vect p}\simeq 0$. We get, as secondary
constraint, $ \frac{{\vect p}^2}{2\,m}- {\vect q}^2 V'({\vect
q}^2)\simeq 0$. For a general potential $V$ this gives a new
condition which in its turn must be stabilized again, and so on. We
can easily end up with incompatibility. Nothing of this kind happens
if we choose the constraint as one of the constants of motion of the
theory.

\subsection{Enforcing a constant of motion as a constraint}

Let us work with the same example  as in the previous
section, i.e. $H({\vect q},{\vect p}) = \frac{{\vect p}^2}{2\,m} +
V({\vect q}^2)$, but now in $\mathbb{R}^3$,
and with $C ({\vect q},{\vect p}) = \epsilon^{3jk} q^j p^k$.
The latter is nothing but one of the conserved angular
momenta.  With this we get
 $ P^{\  \!\!l}({\vect q},\dot{\vect q},\lambda) = m (\dot q^l-\lambda\epsilon^{3jl}q^j)$.
Insertion of  $\vect{P}({\vect q},\dot{\vect q},\lambda)$ into
the constraint $C$ determines
\begin{eqnarray}
\Lambda({\vect q},\dot{\vect q}) \ = \ \frac{\epsilon^{3jk} q^j \dot
q^k}{\alpha}\, ,
\end{eqnarray}
with $\alpha:= (q^1)^2 + (q^2)^2$. Upon evaluation and
elimination of $p$ and $\lambda$ we obtain from the extended
Lagrangian $L_{\rm e}$ the new Lagrangian
\beq L({\vect q}\,\dot{\vect q}) \ = \ \frac{m}{2} \!
\left[\dot{\vect q}^2  -  \frac{(\epsilon^{3jk} q^j \dot
q^k)^2}{\alpha}\right]  \ - \ V({\vect q}^2)\ =  \ \frac{m}{2}
\,\dot{\vect q} \ \!\mathbb{M} \ \! \dot{\vect q} \ - \ V({\vect
q}^2)\, , \label{newl} \eeq
with the projector
\begin{eqnarray} \mathbb{M}^{nk} \ = \ \delta^{nk} \ - \
\frac{\epsilon^{3mn}\epsilon^{3jk} q^m q^j}{\alpha}\, .
\end{eqnarray}
It is easy to check that the projector
$\mathbb{M}$ has $v^k := \epsilon^{3jk} q^j$ as the null vector.

Now we work with the Lagrangian (\ref{newl}). The Lagrangian momenta
are $\hat{\vect p} = {\partial L}/{\partial \dot{\vect q}} =
\mathbb{M}\dot{\vect q}$. The canonical Hamiltonian becomes again
$\frac{{\vect p}^2}{2\,m} + V({\vect q}^2)$ but there is the primary
constraint $\epsilon^{3jk} q^j p^k$ which is now derived from the
definition of the canonical momenta and the use of the null vector
for $\mathbb{M}$. Thus the Dirac Hamiltonian is $H_{\!D}({\vect
q},{\vect p}):= H({\vect q},{\vect p}) + \eta \epsilon^{3jk} q^j
p^k$. Stabilization of this constraint is trivial and there are no
secondary constraints in phase space. In agreement with this fact,
one can check that the Lagrangian (\ref{newl}) does not yield
constraints in tangent (i.e., configuration-velocity)
space.

One can identify the gauge transformation for $L$ as
$\delta_L q^i = \epsilon(t) \{q^i,\,C\}|_{{\vect p}\to
{\vect P},\ \lambda\to \Lambda} = - \epsilon(t)\epsilon^{3ij}q^j$.
It is more instructive to read it by taking cylindrical coordinates
$z,\rho,\theta$; then $\delta_L z=0,\
\delta_L \rho = 0,\ \delta_L\theta = \epsilon$. In
these coordinates the Lagrangian (\ref{newl}) is
\begin{eqnarray}
L \ = \ \frac{1}{2}m\ \!(\dot z^2 \ + \ \dot \rho^2 ) \ - \ V(z^2 \ + \ \rho^2)\,.
\end{eqnarray}
Now the gauge symmetry becomes obvious because there is no
dependence on the angular variable in the Lagrangian. Indeed the
variable $\theta$ is purely gauge. The original, non-gauge,
Lagrangian was $L_{\rm ng} = \frac{1}{2}m(\dot z^2 + \dot \rho^2 +
\rho^2 \dot \theta^2)- V(z^2 +\rho^2)\,$, so we see that the whole
procedure boils down to getting rid of the piece $\rho^2 \dot
\theta^2$. This term was invariant under rigid translations for the
variable $\theta$, that is, rigid rotations around the $z$ axis. The
disappearance of this term makes these rotations a gauge symmetry.

An illuminating consideration can be drawn from this example. At
first sight it could come as a surprise that the implementation of
the constraint, which requires the vanishing of the ``angular
momentum" along the $z$ axis, allows for motions whose projection to
the $x$-$y$ plane has arbitrary dependence in the
variable $\theta$. The correct way of looking at it is the other way
around: in promoting the constant of motion $\epsilon^{3jk} q^j p^k$
to be a constraint, we are also promoting it from being a rigid
symmetry generator to a gauge generator; consequently, the rotations
around the $z$ axis are promoted to gauge transformations. In group
theoretical terms, the implementation of $\epsilon^{3jk} q^j p^k$ as
a constraint has the consequence that a subgroup of the original
rigid symmetry $SO(3)$ gets gauged, precisely that of the rotations
around the $z$ axis.

\subsection{ Relativistic  massive spinless particle}\label{relpart}

Consider the Lagrangian (spacetime indices will be mostly supressed) $L_{\rm
ng}= \frac{1}{2} m\, \dot x^2$ in Minkowski spacetime
with $\eta_{\mu\nu} =  \mbox{diag}(1,-1,\cdots,-1)$,
and the rest mass $m$. Its associated Hamiltonian is $H=
\frac{1}{2\,m}\ \! p^2$. All the momenta are constants of motion, so
we can try to implement them as constraints. We then get the
extended Lagrangian
\begin{eqnarray}
L_{\rm e} \ = \  p\dot x \ - \
\frac{1}{2\,m}\ \! p^2 \ - \ \lambda(p-a)\, ,
\end{eqnarray}
where  in component
notation  $\lambda(p-a) \equiv
\lambda^\mu(p_\mu-a_\mu)$, and $a_\mu$ is a constant $4$-vector. Elimination
of the momenta yields the Lagrangian
\begin{eqnarray}
L_{\lambda}\ = \ \frac{1}{2} m (\dot x-\lambda )^2 \ + \ \lambda\ \!
a\, ,
\end{eqnarray}
 which has the gauge symmetry  $\delta_{\lambda} x^\mu =
\epsilon^{\ \!\!\mu}( \tau),\ \ \delta_{\lambda} \lambda^\mu =
{\dot\epsilon}^{\ \!\!\mu}(\tau)$, with $\epsilon^{\ \!\!\mu}(\tau)$
being arbitrary infinitesimal functions of the
 evolution parameter.  If we further
eliminate the variables $\lambda^\mu$, which have by now acquired
the status of auxiliary variables, we end up with the Lagrangian
\begin{eqnarray}
L\ = \ a\dot x\ - \ \frac{1}{2\,m}\ \! a^2\, .
\end{eqnarray}
The last term is an irrelevant constant. Note  that the EOM for $L$
are void: every trajectory is a solution of the EOM. This conclusion
should not be surprising because all translational symmetries in the
Minkowski target space have been gauged, which results in making any
trajectory acceptable as a solution of the EOM. We have simply
introduced too much gauge freedom.

Instead of trying to gauge the rigid translations in the target
space, we could have decided to gauge the rigid translations along
the world line, that is, the rigid translations in the evolution
parameter. Its associated symmetry in phase space is
$\delta_{\epsilon} x= \epsilon \dot x,\ \ \delta_{\epsilon} p= 0$,
with $\epsilon$ an infinitesimal constant and the generator is the
constant of motion $\frac{1}{2} p^2$. Let us fix the value of this
constant of motion so that $p^2 = m^2$ and require this relation to
become a constraint. This value $p^2 = m^2$ selects trajectories
with unit velocity in Minkowski spacetime, $\dot x^2 = 1$, but after
enforcing this constant of motion as a constraint, a very different
setting emerges, as we will see. For later convenience we consider
the rescaled constant of motion $C= \frac{1}{2\,m} (p^2- m^2)$. In
this case,
\begin{eqnarray}
L_{\rm e} \ = \ p\dot x \ -  \
\frac{1}{2\,m}\ \! p^2- \frac{\lambda}{2\,m}\ \!( p^2- m^2)\, ,
\end{eqnarray}
and elimination of the momenta gives
\begin{eqnarray}
L_{\lambda} \ = \ \frac{m}{2(1\ + \ \lambda)}\ \!\dot x^2\  + \
\frac{1}{2} \lambda \ \! m\, ,\label{33a}
\end{eqnarray}
 which indeed has the gauge symmetry  $\delta_{\lambda} x =
\epsilon(\tau) \frac{\dot x}{1+\lambda},\ \ \delta_{\lambda} \lambda
= \dot \epsilon(\tau)$, obtained under the rules given in
Section~\ref{invert}.  Addition to $L_{\lambda}$ of an irrelevant
constant $m/2$ (which does not affect the dynamics), and a
redefinition $\lambda\to \lambda-1$ allows to write the modified
Lagrangian (for which we keep the same notation) as
\beq L_{\lambda} \ = \ \frac{m}{2\, \lambda}\ \!\dot x^2\  + \
\frac{1}{2} \lambda \ \! m\, ,\label{33b}
\eeq
with gauge transformations $\delta_{\lambda} x= \epsilon({ \tau})
\frac{\dot x}{\lambda},\ \ \delta_{\lambda} \lambda = \dot
\epsilon({ \tau})$. Defining as a new arbitrary function $\xi=
\frac{\epsilon}{\lambda}$, the infinitesimal gauge transformations
read
\begin{eqnarray}
\delta_{\lambda} x \ = \ \xi\dot x,\ \ \ \ \ \ \ \delta_{\lambda}
\lambda \ = \ \frac{\d}{\d \tau}(\xi\lambda)\, , \label{36ab}
\end{eqnarray}
which show directly that $x$ is a scalar and $\lambda$ a scalar
density under the reparametrization $\tau \mapsto \tau-\xi$. The
reader may rightly recognize in $L_{\lambda}$ the familiar
Wheeler--Polyakov's Lagrangian~\cite{Baierlein:62a,Polyakov:87}
\begin{eqnarray}
L_{WP} \ = \ { -}\frac{1}{2} \left({ e}^{-1}(\tau)\ \!
\dot{x}^{\mu}(\tau)\dot{x}_{\mu}(\tau) \  + \ { e}(\tau) \ \!
m^2\right)\, , \label{34a}
\end{eqnarray}
with $\lambda = - m { e} $. The auxiliary variable ${ e}(\tau)$ is
an {\em einbein} (i.e. square-root of the world-line metric) and
$\tau$ is  the world-line parameter (``~label time''). It can be
easily checked that the corresponding action for $L_{\lambda}$ is
invariant under finite reparameterizations of the label time,
 $\tau \ \mapsto  \ \tau' \ = \ f(\tau)$, which, in the
{\em active} view of reparameterization invariance, read
\begin{eqnarray}
 x^{\mu}(\tau) \ \mapsto \ x'^{\mu}(\tau) \ = \
x^{\mu}(f^{-1}(\tau))\,,\ \  \lambda(\tau) \ \mapsto \
\lambda'(\tau)= \left(\frac{\d f^{-1}(\tau)}{\d
\tau}\right)\lambda(f^{-1}(\tau))\, . \label{35a}
\end{eqnarray}
Here $f(\tau)$ is an arbitrary monotonically increasing function of
$\tau$. It is easy to check that the finite
transformations (\ref{35a}) can be obtained from the infinitesimal
transformations (\ref{36ab}) if we set $f(\tau)= \tau - \xi$  and
successively iterate.

The next step is to get rid of the variable $\lambda$  via the
scheme presented in Section~\ref{IIID}. The final Lagrangian $L$
becomes  $L = m\sqrt{\dot x^2}$, which coincides with the usual
square root world-line Lagrangian for relativistic particle. The
corresponding action is  well known to be invariant under
reparameterizations of the label time (i.e. under the first
transformation in (\ref{35a})).
We have thus succeeded in making the original theory
invariant under reparametrizations (or diffeomorphisms). As a bonus
we have recovered the (on mass-shell) equivalence between $L_{WP}$
and the square root world-line Lagrangian.

\section{The gauge principle in relativistic field theory}\label{fields}

\subsection{The minimal setting}\label{minimal}

Let us apply our results to a non-abelian field theory. For
definiteness we will consider a $N$-component complex scalar field
that transforms under the fundamental representation of $SU(N)$
group. The corresponding (non-gauge) Lagrangian density for the free
fields is given by
\beq {\cal L}_{\rm ng} \ = \ \eta^{\mu\nu} (\partial_\mu
{\vect \phi}^*)\cdot(\partial_\nu {\vect \phi}) \ - \ m^2 {\vect
\phi}^*\cdot{\vect \phi}\,. \label{lcomplscal} \eeq
This Lagrangian has clearly $SU(N)$ rigid symmetry
\beq \delta {\vect \phi} \ = \  i \epsilon^a T_a {\vect \phi},\qquad
\delta {\vect \phi}^* \ = \ -i \epsilon^a {\vect \phi}^* T_a\,,
\label {rsymm} \eeq
(Note henceforth that the action of the hermitian matrix
$T_a$ in ${\vect \phi}^* T_a$ undergoes a transposition with
respect to the action of $T_a$ in $ T_a  {\vect \phi}$) with $\epsilon^a$
being infinitesimal constants and
$T_a$ the hermitian $(N\times N)$ matrices spanning a basis of the
Lie algebra of $SU(N)$, $\
[T_a,\,T_b]= i f_ {ab}^c T_c$. To make the rigid
transformation gauge we proceed along the methods
outlined in Sections \ref{ngauge} and \ref{invert}. Let us first
move the description in phase space. The Lagrangian
definition of the momenta is
\beq {\vect \pi} \ = \ \partial_0 {\vect \phi}^* ,\qquad {\vect
\pi}^* \ = \
\partial_0 {\vect \phi}\, , \label{mom}
\eeq
and the Hamiltonian density becomes
\beq {\cal H} \ = \ {\vect \pi}^*\cdot{\vect \pi} \ + \ (\nabla_{\!
i}{\vect \phi}^*)\cdot(\nabla_{\! i}{\vect \phi})\ + \ m^2 {\vect
\phi}^*\cdot{\vect \phi}\, . \label{hdens} \eeq
The constants of motion which generate the rigid $SU(N)$ symmetry
are obtained as coefficients of the infinitesimal constants
$\epsilon^a$ in the space integration of the time component of the
conserved current, which is computed by standard Noether methods
(see, e.g.~\cite{Sundermeyer:1982gv}). We get
\beq j^{~\!\!0}(x) \ = \ i \epsilon^a [{\vect \pi}(x)\cdot~T_a
{\vect \phi}(x) - {\vect \phi}^*(x)T_a\cdot~{\vect \pi}^*(x)]\, .
\label{j0} \eeq
The generator $G = \epsilon^a G_a := \int d^3{\vect x} \ \!j^0(x)$
indeed generates (\ref{rsymm}) together with
\beq \delta {\vect \pi} \ = \ - i \epsilon^a {\vect \pi} T_a ,\qquad
\delta {\vect \pi}^* \ = \ i \epsilon^a {\vect \phi}^* T_a\, .
\label{rsymmp} \eeq
These transformations are in full agreement with the definition of the
Lagrangian momenta (\ref{mom}). The algebra of the generators
\begin{eqnarray} G_a \ = \ i \!\int d^3{\vect x} \ \! \left[{\vect \pi}(x)\cdot~T_a
{\vect \phi}(x)\ - \ {\vect \phi}^*(x)T_a\cdot \ \!{\vect \pi}^*(x)\right]\, ,
\label{ga}
\end{eqnarray}
is $\ \{G_a,\,G_b\}= -f_ {ab}^c G_c$. The opposite  sign
in front of the structure constant $f_ {ab}^c$ is a direct
consequence of the conventional choice $[T_a,\,T_b]= i f_ {ab}^c
T_c$. Contact with our results from Section~\ref{ngauge} can be
established by taking $c_ {ab}^c  = - f_ {ab}^c$.

The extended Lagrangian now takes the form
\begin{eqnarray}
L_{\rm e} \ = \ \int d^3{\vect x} \ \!{\cal L}_{\rm e} \ &=& \ \int
d^3{\vect x}\ \! \left( {\vect \pi}\cdot\dot{\vect \phi}\ + \
\dot{\vect \phi}^*\cdot{\vect \pi}^* \ - \ {\vect \pi}^*\cdot{\vect
\pi} \ - \ (\nabla_{\! i}{\vect \phi}^*)\cdot(\nabla_{\! i}{\vect
\phi})\right. \nonumber \\[1mm]
&&\ - \ \left.m^2 {\vect \phi}^*\cdot{\vect \phi}\ - \ i \lambda^a
({\vect \pi}\cdot~T_a{\vect \phi}\  - \ {\vect
\phi}^*T_a\cdot~{\vect \pi}^* )\right)\,. \label{lefield}
\end{eqnarray}
The gauge transformations for ${\cal L}_{\rm e}$ are given by
(\ref{rsymm}) and (\ref{rsymmp}), but with $\epsilon^a$ now being an
arbitrary infinitesimal function of time, together with the
analogous of (\ref{deltalambda})
\beq \delta\lambda^a(x)\ := \
\partial_0 \epsilon^a(t) \ - \ f_{bc}^a\epsilon^b(t) \lambda^c(x)  \ = \ (D_0\epsilon(t))^a \,.
\label{deltalambdafield} \eeq

Next we proceed as in Section~\ref{invert} to construct the
Lagrangian ${\cal L}_{\lambda}$. We obtain, after some simple
computations
\beq{\cal L}_{\lambda} \ = \ (D_0 {\vect \phi})^*(D_0 {\vect \phi})
\ - \ (\nabla_{\! i}{\vect \phi}^*)\cdot(\nabla_{\! i}{\vect \phi})\
-\ m^2 {\vect \phi}^*\cdot{\vect \phi}\,, \label{almostl} \eeq
with
\beq D_0{\vect \phi}\ := \ \partial_0 {\vect \phi} \ - \ i\lambda^a
T_a {\vect \phi},\qquad (D_0 {\vect \phi})^*\ := \ \partial_0 {\vect
\phi}^* \ + \ i\lambda^a {\vect \phi}^* T_a\,, \label{covd} \eeq
being the usual  gauge covariant derivatives with the
standard covariance condition
$\delta (D_0 {\vect \phi}) = i \epsilon^a(x) T_a D_0 {\vect
\phi}$.

\subsection{Finishing the job}\label{finish}

We have succeeded with ${\cal L}_{\lambda}$ in implementing gauge
invariance in a restricted form. In fact, we have implemented it in
the most minimal way, by adding as many new fields -- the old
Lagrange multipliers -- as dimensions of the original
rigid group we have gauged, and by restricting the infinitesimal
parameters $\epsilon^a(t)$ of the gauge transformation to be only
functions of time, albeit arbitrary. On the other hand,
the above implementation was so minimal that we have lost a big
chunk of the Poincar\' e invariance along the way. Looking at the
structure of the term $D_0 {\vect \phi}$ it is clear that if
Poincar\'e transformations are to be implemented in their entirety,
the fields $\lambda^a$ are nothing else than the time components
$A^a_0$ of vector fields $A^a_\mu$, as $\partial_0 {\vect \phi}$ are
time components of the vector fields $\partial_\mu {\vect \phi}$.
Now we can in a single stroke restore full Poincar\'e invariance and
also let the gauge parameters to have arbitrary dependence on all
the space-time coordinates. We just need to mimic what has been
done for the time coordinate for all the space coordinates. In this
way, gauge invariance is trivially preserved and we recover
Poincar\'e invariance. Then the term $\partial_i {\vect \phi}$ in
the Lagrangian (\ref{almostl}) must be modified to $D_i {\vect \phi}
:=
\partial_i {\vect \phi} - iA^a_i T_a  {\vect \phi}$ and similarly for
$\partial_i {\vect \phi}^*$. The gauge transformations for the gauge
fields will be the generalization of (\ref{deltalambdafield}), namely
$\delta A^a_\mu(x):=
\partial_\mu \epsilon^a(x) - f_{bc}^a\epsilon^b(x) A^c_\mu(x)\,.$
All in all we end up with the well known Lagrangian
\beq {\cal L} = \eta^{\mu\nu}(D_\mu {\vect \phi}^*)\cdot(D_\nu
{\vect \phi})-m^2 {\vect \phi}^*\cdot{\vect \phi}\, , \label{goodl}
\eeq
which is the Lagrangian for the minimal coupling of the complex
scalar fields with the gauge field.

\subsection{The direct way: De~Donder--Weyl formalism \label{dd}}

The way of finishing the job in the previous subsection leaves us
with the uneasiness of having done it with some artifice. The
problem is that the standard canonical formalism destroys the
explicit Lorentz invariance and the procedure in subsection
\ref{minimal} ends up with truly destroying Lorentz invariance,
which then must be restored ``by hand", as done in subsection
\ref{finish}. Fortunately there is a better way.
De~Donder--Weyl formalism~\cite{dedonder}, which preserves manifest
Lorentz invariance in phase space, is a more suited tool to do the
job. Let us go back to the Lagrangian (\ref{lcomplscal}) and define
the Lorentz $4$-component momenta (polymomenta) by
\beq {\vect \pi}^\mu \ = \ \frac{\partial{\cal
L}}{\partial_\mu{\vect \phi}} =
\partial^\mu {\vect \phi}^* ,\qquad {\vect \pi}^{*\mu}\ = \
\frac{\partial{\cal L}}{\partial_\mu{\vect \phi}^*}\ = \
\partial^\mu {\vect \phi}\,. \label{4mom} \eeq
The Hamiltonian, defined in the De~Donder--Weyl formalism
(DWF) through ${\vect \pi}^\mu \cdot \partial_\mu{\vect
\phi} +
{\vect \pi}^{*\ \!\mu}\cdot \partial_\mu{\vect \phi}^* - {\cal L}_{\rm ng}$, becomes
\beq  {\cal H}_{\rm DW} \ = \ {\vect \pi}^\mu \cdot{\vect
\pi}^{*\ \!\nu}\eta_{\mu\nu} + m^2 {\vect \phi}^*\cdot{\vect \phi}\,.
\label{hd} \eeq
To write the extended Lagrangian we will use all four components of
the $SU(N)$ conserved currents, $j^\mu_a=i ({\vect \pi}^\mu T_a\cdot
{\vect \phi} - {\vect \phi}^* T_a\cdot {\vect \pi}^{*\mu})$. This is the
natural way in DWF to maintain a manifest Lorentz
invariance~\cite{foot8}. The associated multipliers $A_\mu^a$ are
then Lorentz $4$-vectors. De~Donder--Weyl's extended Lagrangian can
be then written as
\beq  {\cal L}_{\rm e} \ = \  ({\vect \pi}^\mu)\cdot\partial_\mu{\vect
\phi} \ + \
({\vect \pi}^{\mu})^* \cdot \partial_\mu{\vect \phi}^* \ - \
{\cal H}_{\rm DW} \ - \ i A_\mu^a ({\vect \pi}^\mu \cdot T_a  {\vect
\phi} - {\vect \phi}^* T_a\cdot {\vect \pi}^{*\mu})\,. \label{lefielddd}
\eeq
Finally, applying the methods introduced in Section~\ref{invert}, we
can successively construct Lagrangians ${\cal
L}_{\lambda}$ and ${\cal L}$. By calling the latter as
${\cal L}_{\rm DW}$ we obtain
\beq  {\cal L}_{\rm DW} \ = \ \eta^{\mu\nu}(D_\mu {\vect
\phi}^*)\cdot (D_\nu {\vect \phi})\ - \ m^2 {\vect
\phi}^*\cdot{\vect \phi}\, , \label{goodldd} \eeq
with the covariant derivatives as defined above; $D_\mu {\vect \phi}
:=
\partial_\mu {\vect \phi} - i A^a_\mu T_a  {\vect \phi}$, etc.
By finding ${\cal L}_{\rm DW}$ we have gained a new conceptual access
to gauge field theories in flat space-time.

From here on, the rest is straightforward. One can find the
curvature $[D_\mu,\,D_\nu]$, which transforms under the adjoint
representation of the gauge group and allows for a simple
construction of a gauge invariant Lagrangian with kinetic terms for
the Yang--Mills gauge fields --- and a bonus of new couplings in the
non abelian case. With covariant derivatives and curvatures at
 one's disposal one can analogously formulate other gauge
field theories such as Chern--Simons gauge theory or BF gauge
theory~\cite{Baez:96}.
Non-local gauge invariants like Wilson loops or effective gluon
 masses~\cite{Capri:07} are also at hand.

We have worked out the case of $N$-component complex scalar field
transforming under the $SU(N)$ fundamental representation but we
could have done the same, e.g. for the real-valued field multiplet
in the $SO(N)$ fundamental representation and for the spinorial case
(e.g., for  Dirac or Rarita--Schwinger  fields). Note that the
abelian case is recovered just as a particular case, as it should
be.

\section{World sheet general covariance: the Nambu--Goto closed string\label{NG}}

As another relevant example, we consider the  non-gauge Lagrangian
\beq {\cal L}_{ \rm ng} \ = \  \frac{{{ \rm T}}}{2}
  h^{ab} \partial_a  x^\mu \, \!
 \partial_b  x^\nu \eta_{\mu\nu} \ := \  \frac{ \rm T}{2}
 h^{ab} \partial_a x \ \!  \partial_b x\,,
\label{startl} \eeq
with  the world-sheet metric $h_{ab} =
{\mbox{diag}(1,-1)}$ and the  target-space (or
background) metric $\eta_{\mu\nu}= {\mbox{diag}(1,-1,\ldots, -1)}$.
 ${ \rm T}$ is the string tension. For simplicity we will in the following
work with natural units where  ${ \rm T}=1$.
The target-space functions $x^{\mu}(\tau, \sigma)$
describe the spacetime embedding of the world sheet. In the
following we will suppress the target-space indices. Our aim now is
to gauge the world-sheet rigid translational symmetry
\beq\delta_{\epsilon} x \ = \   \epsilon^a
\partial_a x\,. \label{ngdeltax} \eeq

To prevent any conflicting issue concerning the ``spatial''
($\sigma$) boundary conditions we will deal exclusively in this
section with the closed string. Following Section~\ref{dd}, the
De~Donder--Weyl polymomenta are $p^a = \frac{\partial{\cal
L}}{\partial(\partial_a x)}= h^{ab}\partial_b x$, and the
corresponding De~Donder--Weyl Hamiltonian becomes
\begin{eqnarray}
{\cal H}_{\rm DW}\  = \ p^a \ \!\partial_a x  \ - \  {\cal L}_{ \rm ng} \ = \
\frac{1}{2}h_{ab}\,p^a p^b \, .
\end{eqnarray}
The Noether
conserved current associated with the symmetry (\ref{ngdeltax}) is
found by ordinary methods to be
\begin{eqnarray}
J^a \ =  \ \epsilon^b\!\left(p^a h_{bc}\ \! p^c \ - \
\frac{1}{2}\ \!\delta^a_b p^d
h_{dc}\ \! p^c\right)\,.
\end{eqnarray}
In addition to $\delta_{\epsilon} x$, we need
also to know $\delta_{\epsilon} p^a$. To
compute it we resort momentarily to the standard canonical formalism
and proceed as follow. The world sheet $\tau$-component
of the current is
\begin{eqnarray}
 J^0 \ = \  \epsilon^b\!\left(p^0 h_{bc}\ \! p^c \ - \
\frac{1}{2}\ \!\delta^0_b p^d
h_{dc}\ \! p^c\right)\,,
\end{eqnarray}
 where  $p^1 = h^{11}\partial_1 x = - x'$, so
 $J^0$  has the explicit form
\begin{eqnarray}
J^0 \ = \ \epsilon^0\!\left((p^0)^2 -\frac{1}{2}\
\![(p^0)^2- (x')^2]\right) + \epsilon^1(p^0 x') \ = \
\frac{\epsilon^0}{2} \ \![(p^0)^2+ (x')^2]+ \epsilon^1(p^0 x')\, .
\end{eqnarray}
From this expression the transformations of $p^0$
mediated by the corresponding Noether charge read
\begin{eqnarray}
\delta_{\epsilon} p^0 \ = \  \!\int \!\d\sigma'\left\{p^0(\tau, \sigma),
J^0(\tau, \sigma')\right\} \ = \
\partial_1(\epsilon^0 x' + \epsilon^1p^0)\, . \label{61aa}
\end{eqnarray}
In deriving (\ref{61aa}) we have { allowed} for $\epsilon^a$ to be
an arbitrary infinitesimal  world-sheet function to prepare the
formalism for the gauge transformations we want to implement.

By rewriting $\delta_{\epsilon} p^0$  with the help of
 De~Donder--Weyls' polymomenta  we get
$\delta_{\epsilon} p^0=\partial_1 (\epsilon^1p^0-
\epsilon^0 p^1 )$. Since in the  DWF all
polymomenta play the same role, we infer that the general
transformation law for  $p^a$ is
\beq \delta_{\epsilon} p^a \ = \ \partial_b (\epsilon^b
p^a  - \epsilon^a p^b )\, . \label{ngdeltapddw} \eeq
This should be coupled together with transformations
(\ref{ngdeltax}) which in terms of the De~Donder--Weyl variables
read
\beq\delta_{\epsilon} x \ = \  \epsilon^a h_{ab}\ \! p^b\,.
\label{ngdeltaxddw} \eeq
This last transformation also naturally follows from our
definition of variations $\delta_{\epsilon}$ (cf.
Eq.(\ref{deltapq})), namely
\begin{eqnarray}
\delta_{\epsilon} x \ = \  \!\int
\!\d\sigma'\left\{x(\tau, \sigma), J^0(\tau, \sigma')\right\} \ = \
\epsilon^a h_{ab}\ \! p^b \, , \label{64ab}
\end{eqnarray}
as it, of course, should.

Next, in order to proceed with our program, we define the extended
Lagrangian with Lagrange multipliers $A_{ab}$.  By remembering that
target-space indices are suppressed we obtain
\begin{eqnarray}
{\cal L}_{\rm e} \ &=& \  p^a\partial_a x \ - \ {\cal H}_{\rm DW}\  - \
A_{ab}\!\left(p^a p^b - \frac{1}{2} h^{ab}
 h_{dc}\ \! p^d p^c\right)\nonumber \\
&=& \ p^a\partial_a x
\ - \  \frac{1}{2}\ \! h_{ab}\,p^a p^b \ - \
\frac{1}{2} \ \! B_{ab}\,p^a p^b\,, \label{ngextendedl}
\end{eqnarray}
where $\frac{1}{2}B_{ab}:= A_{ab}-
\frac{1}{2}h_{ab}A_{cd}h^{cd}$ is
symmetric and traceless. This shows that although we initially had
three free Lagrange multipliers  ($A_{ab}$ is symmetric) we
end up with only two, because of the particular structure of the
current $J^a$  and the dimensionality of the world sheet.

Notice the important fact that the new EOM for ${\cal L}_{\rm e} $
imply $\partial_a x= (h_{ab}+ B_{ab}) p^b$, and therefore expression
(\ref{ngdeltaxddw}), originated from (\ref{ngdeltax}) before the
implementation of the Lagrange multipliers, needs to be reformulated
to $\delta_{\epsilon} x = \epsilon^a (h_{ab}+ B_{ab}) p^b$. In turn
this means (cf. Eq.(\ref{64ab})) that the conserved current needs to
be reformulated. It should be noticed that a redefinition
of currents has not been requisite in the previously discussed
systems (apart from relativistic particle in Section~\ref{relpart})
because the Noether currents --- coming from rigid (target-space)
symmetries, do not change when the constraints are imposed. In
contrast, here we deal with currents that come from rigid
world-sheet symmetries and these are influenced when we change
${\mathcal{L}}_{\rm ng}$ to ${\mathcal{L}}_{\rm e}$. Clearly, the
same scenario occurs also for relativistic particle discussed in
Section~\ref{relpart}, but there the change from $\delta_{\epsilon}
x  = \epsilon p/m$ to $\delta_{\epsilon} x  = \epsilon p(1 +
\lambda)/m$ can be assimilated into a redefinition of $\epsilon$
without any extra consequences. This is not the case here (see our
discussion later on).
It is also important to observe that $\delta_{\epsilon} p^a$
as defined by Eq.(\ref{ngdeltapddw}) is not altered
because the metric tensor does not appear in expression
(\ref{ngdeltapddw})  and
one can check that the changes in the current are
exactly absorbed, as regards the computation of $\delta_{\epsilon} p^0$,
with the redefinition
of the relation between $\partial_a x$ and $p^b$, already mentioned.
The above outlined redefinition of the conserved current is just
the first step in an iteration process, with the aim of
consistency, of which we know that at every step the current will be
quadratic in the momenta. Thus this process will result in a final
extended Lagrangian of the general form
\beq {\cal L}_{\rm f} \ = \ p^a\partial_a x - \frac{1}{2}\
\! C_{ab}\,p^a p^b\, , \label{ngextendedlf} \eeq
where $C_{ab}$, which we take symmetric, contains all the
information about the Lagrange multipliers. Seen in retrospect,
(\ref{ngextendedl}) should be interpreted as the first order
expansion of $C_{ab}$ around the world sheet Minkowski metric, so
that $C_{ab} = h_{ab}+ B_{ab}$, with the coefficients $B_{ab}$
now taken infinitesimal. Once this observation is taken into
account, we note that the tracelessness condition for $B_{ab}$
amounts to the condition $\det{ C_{ab}}=-1$ for this $C_{ab} =
h_{ab}+ B_{ab}$. Thus $\det{ C_{ab}}=-1$ is valid at first order
around $h_{ab}$. Repeated iterations of the infinitesimal change
$h_{ab}\mapsto h_{ab}+ B_{ab}$ will be expected to preserve this
condition (cf. Subsection \ref{more}). Thus we end up
with the result that the final extended Lagrangian is supplemented
by the condition
\beq \det{ C_{ab}}\ = \ -1\,. \label{detc} \eeq
The consequences of (\ref{detc}) will be explored later on, in the
next subsection.

If our inputs are correct, the Lagrangian (\ref{ngextendedlf})
should exhibit gauge freedom under the transformations
(with $\epsilon^a $ arbitrary infinitesimal functions),
\beq \delta_{ \epsilon} x \ = \ \epsilon^\alpha C_{ab}\, p^b\,,
\;\;\;\;\;\;\;\; \delta_{\epsilon} p^a \ = \ \partial_b
(\epsilon^b p^a  - \epsilon^a p^b )\, ,\label{ngdeltaxddwf} \eeq
and a
certain (so far unknown) transformation $\delta_{\epsilon} C_{ab}$. This means that
$\delta_{\epsilon} C_{ab}$ should be such that together with
(\ref{ngdeltaxddwf}) it should leave the Lagrangian ${\cal L}_{\rm
ef}$ quasi-invariant, i.e,, with $\delta_{ \epsilon}{\cal L}_{\rm f} $ being a divergence. Let us
now prove the consistency of our scheme by providing  the
explicit form for $\delta_{ \epsilon} C_{ab}$. To this end we first write
\begin{eqnarray}
\delta_{ \epsilon}{\cal L}_{\rm f} \ = \ (\delta_{ \epsilon} p^a)\partial_a x \ + \ p^a
\partial_a (\delta_{ \epsilon} x) \ - \ C_{ab} (\delta_{ \epsilon} p^a) p^b \ -
\ \frac{1}{2}(\delta_{ \epsilon}
C_{ab})p^a p^b\, ,
\end{eqnarray}
and notice that the first term is already a divergence because
\begin{eqnarray}
(\delta_{ \epsilon} p^a)\partial_a x \ = \ \partial_b (\epsilon^b p^a -
\epsilon^a p^b)\partial_a x \ = \
\partial_b [(\epsilon^b p^a - \epsilon^a p^b)
\partial_a x]\, .
\end{eqnarray}
Thus ((div.) stands for divergences),
\bea \delta_{ \epsilon}{\cal L}_{\rm f} &=& ({\rm div.}) + p^a\partial_a
(\epsilon^c C_{cb}\, p^b) - C_{ab} (\delta_{ \epsilon} p^a) p^b
- \frac{1}{2}(\delta_{ \epsilon} C_{ab})p^a p^b \nonumber\\
&=& ({\rm div.})+ p^a\partial_a (\epsilon^c C_{cb}) p^b +
p^a\epsilon^c C_{cb}(\partial_a p^b)- C_{ab}\Big(\partial_c
(\epsilon^c p^a- \epsilon^a p^c)\Big) p^b- \frac{1}{2}(\delta_{ \epsilon}
C_{ab})p^a p^b
\nonumber\\
&=& ({\rm div.})+  p^a\partial_a (\epsilon^c C_{cb}) p^b +
p^a\epsilon^c C_{cb}(\partial_a p^b) +(\epsilon^c p^a- \epsilon^a
p^c)
\partial_c(C_{ab}\,p^b) -
\frac{1}{2}(\delta_{ \epsilon} C_{ab})p^a p^b
\nonumber\\[1mm]
&=& ({\rm div.})+  p^a\partial_a (\epsilon^c C_{cb}) p^b +
p^a\epsilon^c C_{cb}(\partial_a p^b)
 + (\epsilon^c p^a- \epsilon^a p^c)(\partial_c C_{ab})p^b
\nonumber\\[1mm]
&& +\ (\epsilon^c p^a- \epsilon^a p^c)C_{ab}(\partial_c p^b) -
\frac{1}{2}(\delta_{ \epsilon} C_{ab})p^a p^b\,. \label{deltalef} \eea
Consider the next to the last term in (\ref{deltalef}),
i.e., $(\epsilon^c p^a- \epsilon^a p^c)C_{ab}(\partial_c p^b)$. The
second piece cancels another term in (\ref{deltalef}), whereas the
first piece can be written as
\begin{eqnarray}
\epsilon^c p^a C_{ab}(\partial_c p^b)\ = \ \frac{1}{2} \epsilon^c
C_{ab}\partial_c(p^a p^b) \ = \ ({\rm div.})-
\frac{1}{2}\partial_c(\epsilon^c C_{ab})p^a p^b \,.
\end{eqnarray}
All in all we end up with
\bea \delta_{ \epsilon}{\cal L}_{\rm f} &=& ({\rm div.})\ + \
\frac{1}{2}p^a\Big(\partial_a (\epsilon^c C_{cb})\ + \
\partial_b (\epsilon^c C_{ca})\Big)p^b
 \ + \ (\epsilon^c p^a- \epsilon^a p^c)(\partial_c
C_{ab})p^b \nonumber\\ && - \ \frac{1}{2}\partial_c(\epsilon^c
C_{ab})p^a p^b -\frac{1}{2}(\delta_{ \epsilon} C_{ab})p^a p^b\nonumber\\[2mm]
&=& ({\rm div.}) \ + \ \frac{1}{2}p^a\Big(\partial_a (\epsilon^c
C_{cb})+\partial_b (\epsilon^c C_{ca})\Big)p^b \ + \
p^a\epsilon^c(\partial_c C_{ab})p^b\nonumber \\
&& - \ \frac{1}{2}p^a\Big(\epsilon^c(\partial_a
C_{cb})+\epsilon^c(\partial_b C_{ca})\Big)p^b \ - \
\frac{1}{2}\partial_c(\epsilon^c C_{ab})p^a p^b \ - \
\frac{1}{2}(\delta_{ \epsilon} C_{ab})p^a p^b\, , \label{deltalef2} \eea
which implies that under the transformation
\beq \delta_{ \epsilon} C_{ab} \ = \ \epsilon^c\partial_c C_{ab} \ + \
C_{cb}\partial_a\epsilon^c \ + \ C_{ac}\partial_b\epsilon^c \ - \
C_{ab}\partial_c\epsilon^c\,, \label{tensdenscov} \eeq
the Lagrangian ${\cal L}_{\rm f}$ is indeed quasi-invariant. Note that this solution
(\ref{tensdenscov}) for the transformations rules of $C_{ab}$ is unique. Equation
(\ref{tensdenscov}) is the Lie derivative of a covariant tensor
density $(0,2)$ of weight $-1$ along ${\vect \epsilon}$, i.e., $\delta_{ \epsilon} C_{ab}
= \pounds_{\vect \epsilon} C_{ab}$. Its inverse matrix, which we denote as
$C^{ac}$ will then be a  contravariant tensor density $(2,0)$ of weight $+1$,  which
then transforms according to
\beq \delta_{ \epsilon} C^{ab} \ = \ \epsilon^c\partial_c C^{ab} \ - \
C^{cb}\partial_c\epsilon^a \ - \ C^{ac}\partial_c\epsilon^b \ + \
C^{ab}\partial_c\epsilon^c  \ =  \ \pounds_{\vect \epsilon} C^{ab}\,. \label{tensdenscont} \eeq
The result (\ref{tensdenscov}) is a very good news because the elimination of the
momenta from their own EOM in (\ref{ngextendedl})
produces the Lagrangian ${\cal L}_{\lambda}$ --- which in this context
is more reasonable to denote as ${\cal L}_{\rm C}$ (and similarly substitute $\delta_{\lambda}$ by
$\delta_{_{\rm C}}$) --- which reads
\beq {\cal L}_{{\rm C}}(x, \partial_a x, C_{bc}) \ = \
 {\cal L}_{\rm f}(x, \partial_a x, P^c(x, \partial_a x,
C_{de},), C_{de}) \ = \  \frac{1}{2}C^{ab}\partial_a x\partial_b
x\,. \label{lc} \eeq
The latter is  a scalar density under the transformations
(\ref{ngdeltax}) and
(\ref{tensdenscont}), indeed $\delta_{_{\rm C}} {\cal L}_{\rm C} =
\partial_a(\epsilon^a{\cal L}_{\rm C})$. Because transformations
(\ref{ngdeltax}) and (\ref{tensdenscont}) are respectively Lie
derivatives for scalars and for tensor densities, they --- similarly
as in the general relativity~\cite{Sundermeyer:1982gv} --- express
diffeomorphism invariance (or general covariance) of the theory.

\subsection{The condition $\det{C}=-1$}

The Lagrangian ${\cal L}_{\rm C}$ is not the end of the story
because the auxiliary variables $C_{ab}$ satisfy the
additional condition $\det{C_{ab}}=-1$. First notice that this
condition is compatible with the gauge symmetry because
$\det{C_{ab}}$ behaves as a scalar under the gauge transformation (\ref{tensdenscov}),
\begin{eqnarray}
\delta_{\epsilon} \det C \ = \ \epsilon^a\partial_a(\det{C})\, .
\end{eqnarray}
As a
by-product we see that by requiring the extra constraint  $\det
C_{ab} =-1$ the gauge freedom stays intact.

In practice, one may consider two ways to implement the condition
$\det{C_{ab}}=-1$ into ${\cal L}_{\rm C}$. One possible procedure is
to introduce new gauge freedom by defining
$C_{ab}=\frac{1}{\sqrt{-g}}\,g_{ab}$, with $g_{ab}$ an arbitrary
symmetric tensor in the world sheet of signature $\{+, - \}$, and
$g:= \det{g_{ab}}$ (note that $\det(\frac{1}{\sqrt{-g}}\,g_{ab})=-1$
and $C^{ab} = \sqrt{-g} g^{ab}$). The new gauge freedom is Weyl
invariance, $g_{ab} \mapsto \Lambda(\tau,\sigma) g_{ab}$. This new
gauge freedom compensates for the fact that $g_{ab}$ has three
components whereas $C_{ab}$ had only two. The result is the familiar
 non-linear $\sigma$ model
Lagrangian~\cite{Brink:1976sc,Polyakov:1981rd,foot9} for bosonic
string theory,
\beq {\cal L}_{\sigma} \ = \ \frac{1}{2}\sqrt{-g}
g^{ab}\partial_a x\partial_a x\,. \label{lp} \eeq
It is well known that  at the classical level one can eliminate
$g_{ab}$, which are an auxiliary variables in (\ref{lp}), by
plugging their own EOM into (\ref{lp}). The result is the
Nambu--Goto Lagrangian.  Quantum mechanically is the issue more
delicate. Instead of eliminating $g_{ab}$ via its EOM, one should
perform a Feynman path integral, and use the standard Fadeev--Popov
procedure to deal with the local symmetries and gauge fixing. When
this is done correctly~\cite{Polyakov:87}, one finds that there is a
conformal anomaly unless the target-space dimension is $D = 26$. But
even in $26$ dimensions it is not yet clear whether off mass-shell
fluctuations of the Nambu--Goto  and the non-liner $\sigma$-model
actions contribute in the same way, say into string partition
function. As we are interested here only in classical level
description we will not push  this point further.

The second procedure consists in enforcing $\det{C}=-1$ with a
Lagrange multiplier. One modifies the Lagrangian (\ref{lc}) so that
the new  Lagrangian is
\beq \tilde{\cal L}_{\rm C} \ = \ \frac{1}{2}\ \! C^{ab}\partial_a
x\partial_b x \ + \ \lambda(t-1)\,, \label{llambda} \eeq
where $t:=\sqrt{-\det{C_{ab}}}$ (the square root is
introduced for a technical convenience). Since the first term in
(\ref{llambda}) is already a scalar density, the transformation
properties of the multiplier $\lambda$ must be also those of a
scalar density, i.e. $\delta_{_{\rm C}} \lambda =
\partial_a(\epsilon^a\lambda)$. Using the fact that
$C_{ab}$ have become auxiliary variables for (\ref{llambda}), we
obtain from their own EOM that $C_{ab}=
\frac{1}{\lambda\,t} \partial_a x\partial_b x$, and therefore $t$ is determined
as
\begin{eqnarray}
t \ = \ \frac{1}{\sqrt{\lambda}} (-\det{ \partial_a x \ \! \partial_b
x})^{\frac{1}{4}}\,.
\end{eqnarray}
Plugging this result into (\ref{llambda}) we get
\beq {{\bar{\cal L}}}_{\rm C} \ = \  2
\sqrt{\lambda}(-\det{ \partial_a x \ \! \partial_b
x})^{\frac{1}{4}}-\lambda\,. \label{llambda2} \eeq
Now the multiplier $\lambda$ has turned an auxiliary variable. Its
EOM determines $\lambda = (-\det{ \partial_a x \ \! \partial_b
x})^{\frac{1}{2}}$. Substitution of this result into
(\ref{llambda2}) yields the Nambu--Goto Lagrangian
\beq {\cal L}_{ \rm NG} \ = \ (-\det{ \partial_a x \ \! \partial_b
x})^{\frac{1}{2}}\,. \label{lng}
\eeq
This again reconfirms the fact that on mass-shell ${\mathcal{L}}_{\sigma} \cong{\mathcal{L}}_{\rm NG} $.

\subsection{Further considerations}
\label{more}

There is a strong parallelism between our way of
obtaining the world-sheet general covariance and
the approach~\cite{kraichnanetal} to general relativity out of the
requirement of self consistency of the coupling of the energy
momentum tensor of an initially Minkowskian theory to a massless
spin-$2$ field. The presence of the coupling term produces changes
in the energy momentum tensor which in its turn redefine the
coupling term, making it non linear in the spin-$2$ field. An
interaction procedure is set to work and the final result is the
appearance of the metric tensor field and general covariance. In our
case the Lagrange multipliers $B_{ab}$ play the role of the spin-$2$
field. A self consistency requirement also appears because the
conserved current for world sheet translation invariance has changed
due to the presence of the new term with the multipliers. In fact in
the  DWF we enforce all the components of the current to become
constraints, and thus the Lagrange multipliers $B_{ab}$ are in fact
coupled to the energy momentum tensor. The difference is that in our
case, due to the particular structure of the current, we end up with
a density tensor field $C_{ab}$ of weight $-1$ that must satisfy
$\det{C_{ac}}=-1$.

Let us elaborate a bit more on the requirement $\det{C_{ab}}=-1$. This
condition is crucial for our purposes. In fact we have found the
fulfillment of this condition for configurations of $C_{ab}$ around
the flat spacetime metric and we have checked that the extension of
this result to any configuration is fully compatible with gauge
freedom. We could also argue that since we have found only two
degrees of freedom --- those of traceless symmetric $B_{ab}$ --- around
the flat spacetime metric, to preserve this number we must accept
that the components of $C_{ab}$ are constrained by
 a condition of the type $f(C_{ab}) = {\rm constant}$.
If we make the reasonable
assumption that this condition is geometrical --- since the Lagrangian
(\ref{lng}) already is ---, we conclude that it
should be a scalar under diffeomorphisms. But the only scalar we can
construct out of the components of the tensor density $C_{ab}$ is
just its determinant, and to fix its value we need only to consider
the configurations around $h_{ab}$.

It is remarkable that as a way to perform the covariant quantization
of the bosonic string, Kato and Ogawa~\cite{Kato:1982im} used
essentially the Lagrangian (\ref{llambda}) as a Lagrangian
equivalent to (\ref{lp}). On the other hand,
Siegel~\cite{Siegel:1985xj}, see also~\cite{Fedoruk:2006de}, used
the extended Lagrangian (\ref{ngextendedlf}) with the specific
requirement $\det{C_{ab}}=-1$. In our approach (\ref{ngextendedlf}) and
(\ref{llambda}) are consequences of gauging the world-sheet rigid
translational symmetry of the Lagrangian (\ref{startl}).

Finally let us stress that the dimensionality of the
world sheet plays a crucial role in our derivation of the
Nambu--Goto Lagrangian (\ref{lng}) through gauging
the rigid world-sheet
translational symmetry (\ref{ngdeltax}). It is only when the world
sheet is $2$-dimensional that the Lagrange multipliers are
constrained so as to satisfy an additional condition which eventually
leads to the requirement $\det{C_{ab}}=-1$.

\section{Conclusions \label{concl}}

\begin{figure}
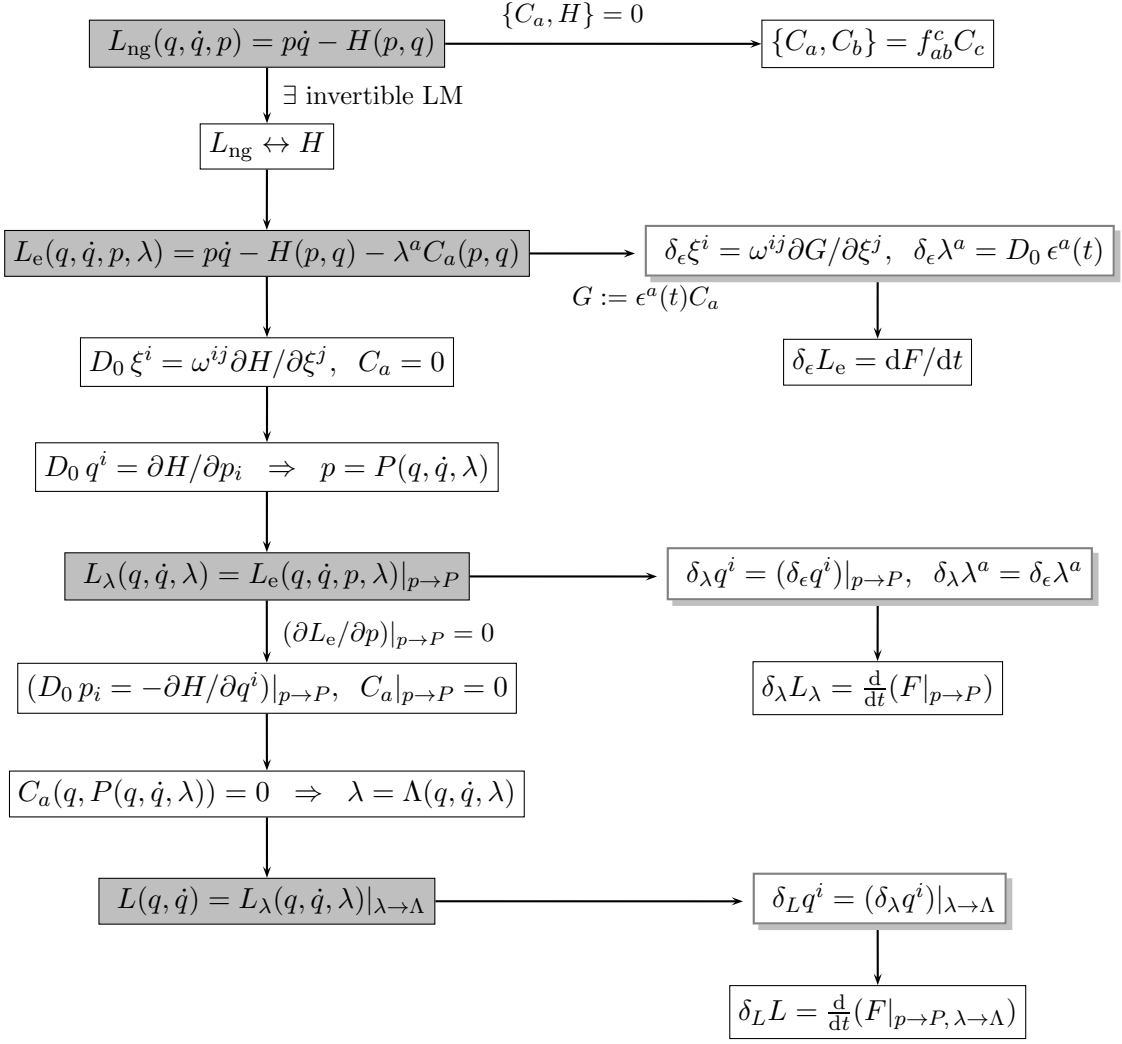

  \centering
  \hspace{1cm} 
  \begin{psmatrix}[mnode=r,colsep=1.4,rowsep=0.7]
   [name=L_n] \ps{ {$L_{\rm ng}(q,\dot{q},p) = p\dot{q} - H(p,q)$}}  &
   [name=C]\pw{{$\{ C_a, C_b \} = f_{ab}^c C_c$}}
   \\[0pt]
   [name=a]  \pw{$L_{\rm ng} \leftrightarrow H$} &
                        \\[0pt]
    [name=L_e]  \ps{{$ L_{\rm e}(q,\dot{q},p, \lambda) = p\dot{q} - H(p,q) - \lambda^{a}
    C_{a}(p,q)$}} &
    [name=c] \rxy{$\delta_{\epsilon}\xi^i = \omega^{ij}
    \partial G/ \partial \xi^j,
    \; \;  \delta_{\epsilon}\lambda^a = D_0\, \epsilon^a(t) $}{\footnotesize{$G : = \epsilon^a(t)C_a$}}
    \\[0pt]
    [name=x] \pw{$D_0\, \xi^i = \omega^{ij}\partial H/\partial \xi^j, \; \; C_a = 0$}
                       &
    [name=g] \pw{$\delta_{\epsilon}L_{\rm e} = \d F/\d t$}
    \\[0pt]
    [name=g00] \pw{$D_0\, q^i  = \partial H/\partial p_i \;\; \Rightarrow \;\;
                       p = P(q,\dot{q}, \lambda)$} &
                       \\[0pt]
    [name=L_l]  \ps{ {$L_{\lambda}(q,\dot{q},\lambda) =
    L_{\rm e}(q,\dot{q},p,\lambda)|_{p\rightarrow P}$}} &
    [name=c1] \rxy{$\delta_{\lambda}q^i =
   (\delta_\epsilon q^i)|_{p\rightarrow P},
    \; \;  \delta_{\lambda}\lambda^a = \delta_{\epsilon}\lambda^a$}{}
    \\[0pt]
    [name=g000] \pw{$(D_0\,p_i = -  \partial H/\partial q^i )|_{p\rightarrow P},
    \;\; C_a|_{p\rightarrow P} = 0$} &
    [name=g1] \pw{$\delta_{\lambda}L_{\lambda} = \frac{\d}{\d t}(F|_{p\rightarrow P})$}
    \\[0pt]
    [name=g10] \pw{$C_a(q, P(q,\dot{q}, \lambda))= 0\;\; \Rightarrow
    \;\; \lambda = \Lambda(q,\dot{q},\lambda)$}\\[0pt]
    [name=L]  \ps{ {$L(q,\dot{q}) =
    L_{\lambda}(q,\dot{q},\lambda)|_{\lambda \rightarrow \Lambda}$}}& [name=c2] \rxy{$\delta_{L}q^i =
   (\delta_\lambda q^i)|_{\lambda\rightarrow \Lambda}$}{} &
   \\[0pt]
                       &    [name=g2] \pw{$\delta_{L}L = \frac{\d}{\d t}(F|_{p\rightarrow P, \, \lambda\rightarrow \Lambda})$}
                        \\[0cm]
    \ncline{->}{L_n}{C}^{\footnotesize{$\{C_a,H\}=0$}}
    \ncline{->}{L_n}{a}>{\footnotesize{$\exists$ invertible LM}}
    \ncline{->}{a}{L_e}
    \ncline{->}{L_e}{c}
    \ncline{->}{c}{g}
    \ncline{->}{L_l}{c1}
    \ncline{->}{c1}{g1}
    \ncline{->}{L_e}{x}
    \ncline{->}{x}{g00}
    \ncline{->}{g00}{L_l}
    \ncline{->}{L}{c2}
    \ncline{->}{c2}{g2}
    \ncangle[angleA=-90,angleB=180]{->}{h}{j}
    \ncangle[angleA=-90,angleB=0]{->}{i}{j}
    \ncangle[angleA=-90,angleB=0]{->}{j}{k}
    \ncangle[angleA=180,angleB=180,arm=100pt]{->}{L_e}{k}
    \ncline{->}{k}{L_l}
    \ncline{->}{L_l}{g000}>{\footnotesize{$(\partial L_{\rm e}/\partial p)|_{p\rightarrow P} = 0$}}
    \ncline{->}{g000}{g10}
    \ncline{->}{g10}{L}
  \end{psmatrix}
  \vspace{0.2cm} \hrule
  \caption{\footnotesize{The sequence diagram summarizing the basic logical steps
leading from $L_{\rm ng}$  to $L$. The abbreviation LM stands for
Legendre map while $F$ denotes some phase-space function which is
linear in $\epsilon$ and its derivatives.}} \label{G.2}
\end{figure}

Let us summarize our findings. Our starting point is a non-gauge
theory, defined by a regular Lagrangian $L_{\rm ng}$. We assume that
in the phase space formulation such a
theory has a Lie algebra of time independent constants of motion.
Next we enforce these constants of motion as first class constraints
by adding them to the Hamiltonian with a set of Lagrange
multipliers. Then we perform the inverse Legendre transformation to
end up with a new (extended) Lagrangian $L_{\rm e}$ whose
configuration space now includes the Lagrange multipliers as new
variables. We then observe that this new theory has gauge symmetries
and that the gauge group is generated by the constraints, as
expected. We also observe that in general the new variables are
auxiliary and that they can be further eliminated from the formalism
by plugging  into the new Lagrangian their determination through
their own equations of motion. This yields the final
gauge invariant Lagrangian $L$. This last step may be
problematic with regard to quantization because the final theory
will in general be of non-polynomial nature.
Another option is to enlarge the theory with the addition
of new gauge invariant terms that make these auxiliary variables dynamical.  The passage
from $L_{\rm ng}$ to $L$, is schematically illustrated in the
sequence diagram in Fig.\ref{G.2}.

In the special case of relativistic field theories we have noticed
that our program is best carried out if the canonical setting  is
taken along the lines of the De~Donder--Weyl approach. Such
formalism is particularly suitable because it keeps manifest Lorentz
invariance from the very scratch.  The simplicity with which this
gauging procedure can be performed within this formalism
is remarkable.

We have illustrated the DWF  by applying it
to the case of $N$-component complex scalar field transforming under
the $SU(N)$ fundamental representation but we could have done the
same, e.g. for the real-valued field multiplet in the $SO(N)$
fundamental representation, for the spinorial case, etc. It
should be, nevertheless, noted that the role of the DWF is purely
instrumental, and that once the Lagrangian for the gauge theory has
been obtained (see, for instance, Eq.(\ref{goodldd})),
one can proceed either with Lagrangian or with standard canonical
methods, without having to rely again on the DWF.

As another relevant example we have derived the Nambu--Goto
Lagrangian for the closed bosonic string by gauging the world sheet
rigid translational symmetry of a non-gauge Lagrangian. Our strategy
has again relied on the DWF and it entailed an iteration procedure
very close in spirit to the approach to Einstein's general theory of
gravitation~\cite{kraichnanetal} in which a consistency argument on
the coupling of a massless spin $2$ field with the total
energy-momentum tensor (including matter fields) yields ultimately
the Einstein--Hilbert action. It should be, however, stressed that
because in our reasonings the dimensionality of the world-sheet has
played a crucial role, it is not yet clear if a similar iterative
procedure can be applied, e.g., to relativistic Dirac--Nambu--Goto
membranes (or {\it p}-branes).

The above considered examples clearly indicate that the
gauge principle, i.e., the gauging of a rigid group of
symmetries, can be alternatively recast in the language
of constrained systems with the gauge fields appearing first as
Lagrange multipliers for the enforcement of the constants of motion
as constraints. The rationale of the procedure is based on the fact
that rigid symmetries are generated by constants
of motion, whereas gauge symmetries by first class constraints.
Thus to gauge a group of rigid symmetries is tantamount to enforce the
generating constants of motion as constraints. Note also that the
role of the gauge fields as multipliers is temporary, because after
the implementation of the inverse Legendre transformation they
typically become auxiliary variables. Finally, when the Lagrangian
is modified with new gauge invariant additions to provide for
kinetic terms for the gauge fields, they become dynamical variables
on their own.

We notice also that the constraints $C_a$, directly originated from
the former constants of motion of the non gauged theory, are primary
constraints, but that does not mean that our framework is limited to
this kind of constraints and can not give rise to secondary
constraints. On the contrary, the examples provided in section
\ref{fields} show that, due to the presence of the kinetic terms for
the gauge fields - which are the former Lagrangian multipliers - in
the final Lagrangian, secondary constraints may arise, as it is
indeed the case for the Yang-Mills gauge theories.

With the benefit of hindsight, we observe that this route of
enforcing constants of motion as constraints could have been taken
from the very beginning as an alternative way to the gauge
principle, because at the time when the Yang--Mills theory was
formulated, the foundations and development of the theory of
constrained systems were already in place.

We believe that the presented formulation can be also conveniently
applied in the 't Hooft program~\cite{thooft} where the extended
Lagrangians (\ref{le}) formulated with the help of constants of
motion have played a pivotal role in construction of emergent
dynamical systems~\cite{blasone:05a, blasone:05b}. This issue would
deserve further investigation.

\section{Acknowledgments}
P.J. is grateful to H.~Kleinert for instigating discussions. J.M.P. thanks J.~Gomis for
useful discussions and for pointing out a relevant reference.
This work was partially supported by the Ministry of Education
of the Czech Republic (research plan no.
MSM 6840770039), and by the Deutsche Forschungsgemeinschaft under
grant Kl256/47. J.M.P. acknowledges partial support form  MCYT FPA 2007-66665, CIRIT GC
2005SGR-00564, Spanish Consolider-Ingenio 2010 Programme CPAN
(CSD2007-00042).



\begin{thebibliography}{99}
\bibitem{foot4} Dirac introduced the
concept of a first class function as a function whose Poisson
bracket with the constraints vanishes on the constraints'
surface. Gauge symmetries
are made up with linear combinations of first class constraints,
which appear as coefficients of an expansion in terms of the
arbitrary functions and their time derivatives up to a certain
order.
\bibitem{foot7} Note that the operator $\frac{\d}{\d t}$ means different
things depending on the variables we consider. For instance, with
variables $q,\ p, \lambda$, we will have $\frac{\d}{\d t} =  \dot
q\frac{\partial}{\partial q} + \dot p\frac{\partial}{\partial p} +
\dot  \lambda\frac{\partial}{\partial \lambda} + \ddot
q\frac{\partial}{\partial \dot q}+ \ldots$. Instead, if we send
$p\to P(q,\dot q,\lambda)$, we will have $\frac{\d}{\d t} =  \dot
q\frac{\partial}{\partial q}  + \dot \lambda\frac{\partial}{\partial
\lambda} + \ddot q\frac{\partial}{\partial \dot q}+\ldots$.
\bibitem{foot8} To strengthen this point note that the
currents associated with gauge symmetries must always vanish on
shell, except for possible divergences of arbitrary antisymmetric
tensors. In the De~Donder--Weyl formalism all the components of the
gauge currents are constraints in the new phase space.
\bibitem{foot9} Actually the name string (non-linear
$\sigma$) model Lagrangian is a little unfair, since the Lagrangian
(\ref{lp}) was discovered independently by several researchers. But
in our view it is a bit clumsy to talk about the
Brink--Di~Vecchia--Howe--Deser--Zumino--Polyakov Lagrangian.
\end{thebibliography}
\end{document}